\documentclass[3p]{elsarticle}
\usepackage{hyperref,graphicx,subfig}
\usepackage{mathrsfs}
\usepackage{xspace}

\usepackage{lineno}
\modulolinenumbers[5]

\journal{Reviews in Physics (REVIP)}

\newcommand{\mt}{\ensuremath{m_{\mathrm{top}}}}
\newcommand{\mtmc}{\ensuremath{m_{\mathrm{top}}^{\mathrm{MC}}}}
\newcommand{\mtpole}{\ensuremath{m_{\mathrm{top}}^{\mathrm{pole}}}}
\newcommand{\mh}{\ensuremath{m_{\mathrm{H}}}}
\newcommand{\ifb}{\ensuremath{\mathrm{fb}^{-1}}}
\newcommand{\met}{\ensuremath{E_\mathrm{T}^\mathrm{miss}}}
\newcommand{\ttbar}{\ensuremath{t\bar t}}
\newcommand{\ppbar}{\ensuremath{p\bar p}}

\newcommand{\ttbarll}{\ensuremath{t\bar t \to\mbox{di-lepton}}\xspace}
\newcommand{\ttbarlj}{\ensuremath{t\bar t \to\mbox{lepton+jets}}\xspace}
\newcommand{\ttbaraj}{\ensuremath{t\bar t \to\mbox{all-jets}}\xspace}
\newcommand{\ttbarmet}{\ensuremath{t\bar t \to E_{\rm T}^{\rm miss}\mbox{+jets}}\xspace}
\begin{document}

\begin{frontmatter}

\title{Top-quark mass measurements: review and perspectives}

\author{Giorgio Cortiana}
\address{Max-Planck-Institut f{\"u}r Physik,
     F{\"o}hringer Ring 6, D-80805 M{\"u}nchen, Germany}

\begin{abstract}
  The top quark is the heaviest elementary particle known and its mass
  (\mt) is a fundamental parameter of the Standard Model (SM). The
  \mt\ value affects theory predictions of particle production
  cross-sections required for exploring Higgs-boson properties and
  searching for New Physics (NP). Its precise determination is
  essential for testing the overall consistency of the SM, to
  constrain NP models, through precision electroweak fits, and has an
  extraordinary impact on the Higgs sector, and on the SM
  extrapolation to high-energies.  The methodologies, the results,
  and the main theoretical and experimental challenges related to the \mt\
  measurements and combinations at the Large Hadron Collider (LHC)
  and at the Tevatron are reviewed and discussed.
  Finally, the prospects for the improvement of the \mt\ precision
  during the upcoming LHC runs are briefly outlined.
\end{abstract}


\end{frontmatter}


\section{Introduction}

\begin{figure}[tbp!]
\centering
\subfloat[]{
  \includegraphics[width=0.55\textwidth]{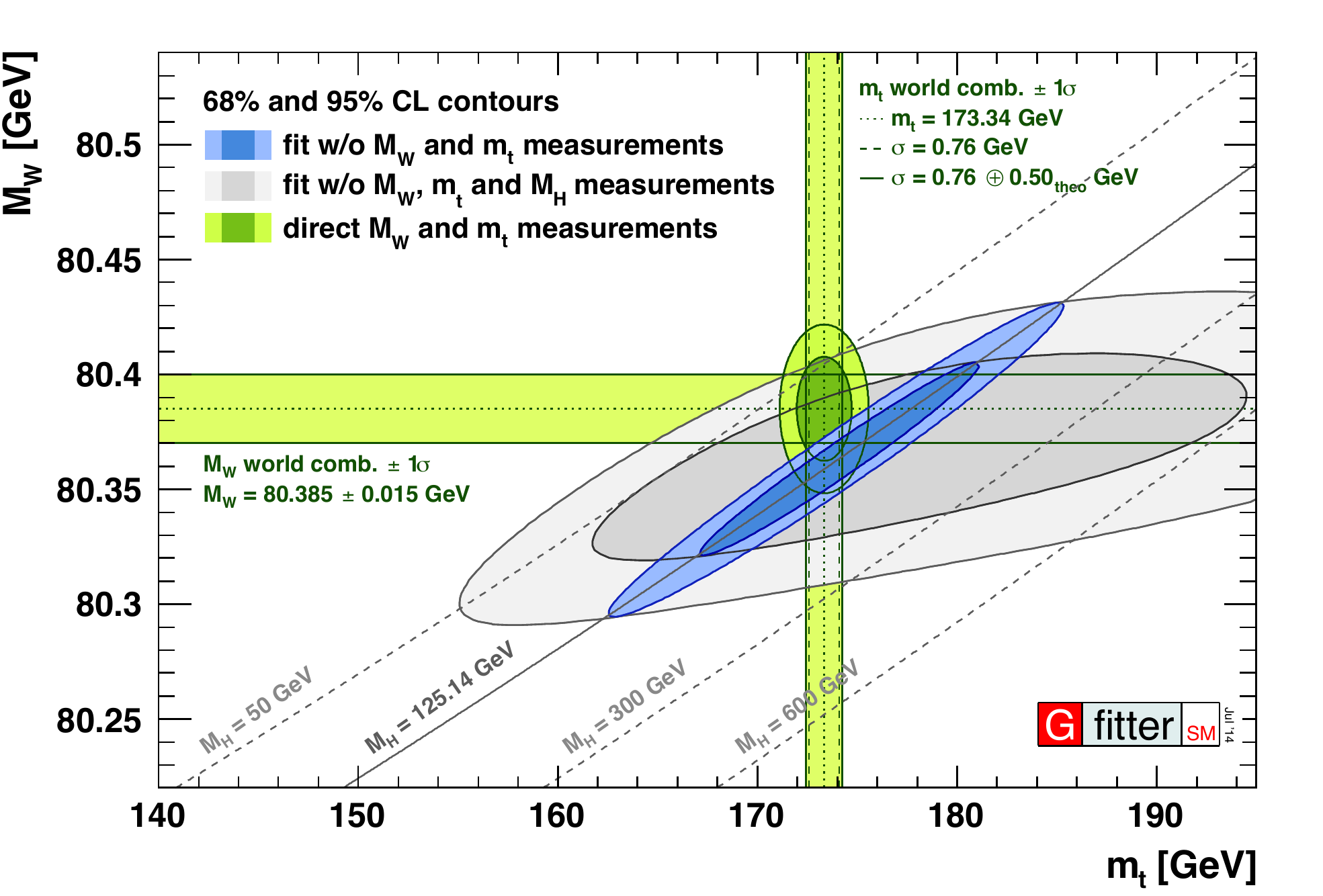}}
\subfloat[]{
  \includegraphics[width=0.396\textwidth]{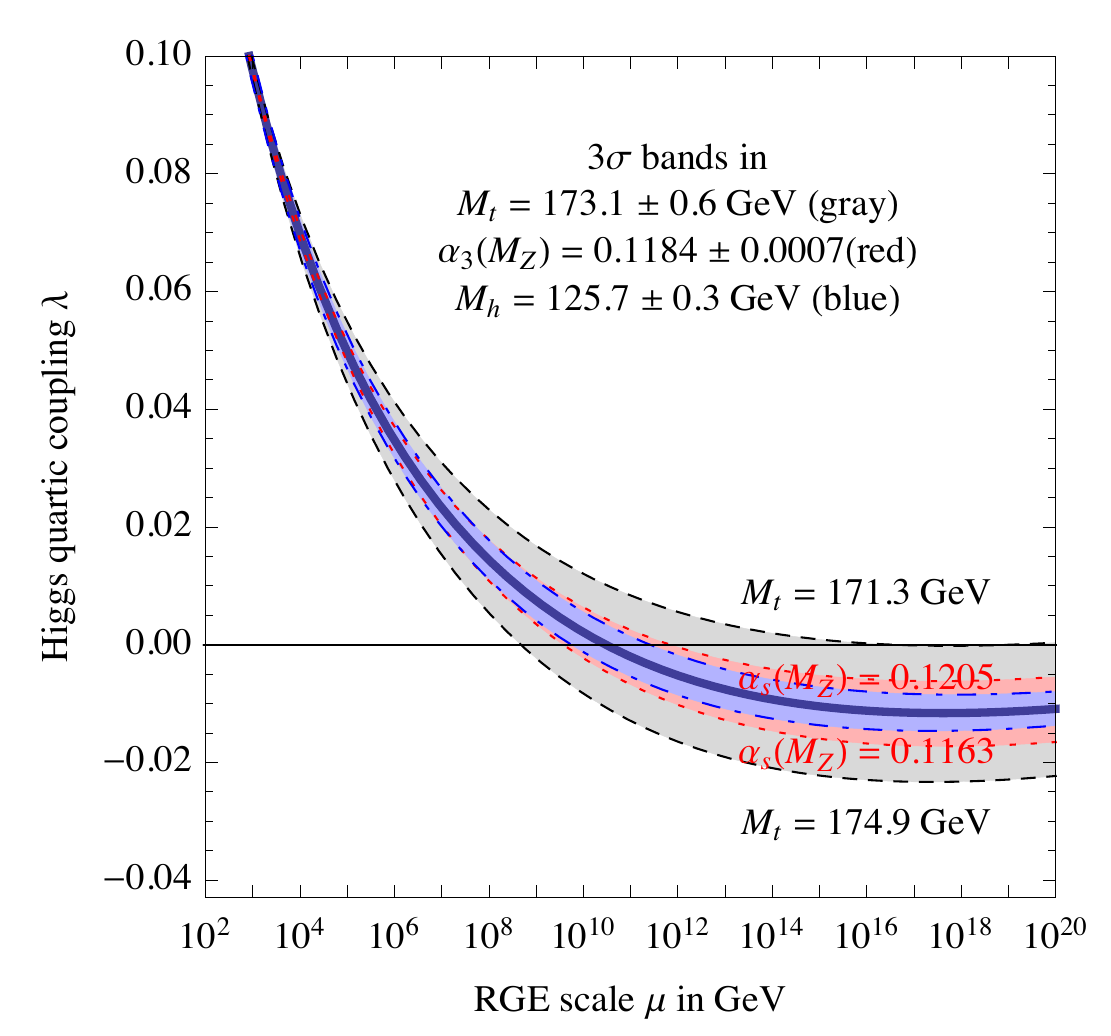}}\hfill
\caption{(a) The 68\% and 95\% CL contours for the indirect
  determination of $m_W$ and \mt\ from
  global SM fits to electroweak precision data (from
  Ref.~\cite{EWFITS}). (b) Evolution of the Higgs-boson self coupling
  $\lambda$, within the renormalisation group equation (RGE), for the
  central values of $\mh = 125.7$~GeV, $\mt = 173.1$~GeV and
  $\alpha_S(m_{Z}) = 0.1184$ (solid curve), and variation of these
  central values by $\pm 3\sigma$ for the blue, grey and red, dashed
  curves, respectively (from Ref.~\cite{Degrassi}).
  \label{fig:intro}}
\end{figure}

Naturally complementing direct searches for new physics (NP) phenomena,
precision measurements of the properties of the fundamental particles
constitute an extremely successful path to refine our knowledge of
high-energy physics and of its implications on the evolution of the
Universe.  
In this context, the top quark plays a special role: its lifetime is
extremely short ($\approx 10^{-25}$~s) and inhibits top-quark bound
states and top-quark flavoured hadrons to be formed, offering a
unique possibility to study the properties of the particle as a
quasi-free quark (see Refs.~\cite{topphysLHC, topphys} for recent
reviews on the subject).
The top quark is the heaviest elementary particle currently known and
its mass (\mt) is a fundamental parameter of quantum chromodynamics
(QCD), the quantum field theory describing the strong interactions of
quarks and gluons. The value of \mt\ affects theory predictions of
particle production cross-sections required for exploring Higgs-boson
properties and searching for NP phenomena. Its precise determination
is essential for testing the overall consistency of the Standard Model
(SM) and to constrain NP models through precision electroweak fits.
Figure~\ref{fig:intro}(a), from Ref.~\cite{EWFITS}, displays the 68\%
and 95\% confidence level (CL) contours for the indirect determination
of the mass of the $W$ boson ($m_W$) and \mt\ from global SM fits to
electroweak precision data.  The blue (grey) areas illustrate the fit
results when including (excluding) the direct Higgs-boson mass
measurements~\cite{ATLASHiggs, CMSHiggs}. The contours are
compared with the direct measurements of $m_W$ and \mt, shown by the
horizontal  and vertical green bands, that are excluded from the fits.

In addition, owing to its large value, of the order of the electroweak
symmetry breaking energy scale, \mt\ has a direct impact on the Higgs
sector of the SM, and on extrapolations of the SM to high-energy
scales~\cite{Degrassi,WhyYTop}.
With the discovery of a Higgs boson~\cite{ATLASHiggs, CMSHiggs} at the
Large Hadron Collider (LHC) with a mass of $\mh = 125.09 \pm
0.24$~GeV~\cite{CombHiggsMass}, precision measurements of the
top-quark mass take a central role in answering the question of the
stability of the electroweak vacuum: top-quark radiative corrections
can drive the Higgs-boson self coupling ($\lambda$) towards negative
values, potentially leading to an unstable vacuum.  The determination
of the energy scale ($\mu$) at which this happens, possibly requiring
new physics at lower or comparable energies, is strongly influenced by
the precision of the top-quark mass measurement and by the
interpretation of \mt\ in a clear theoretical framework ($\mu$ varies
by several orders of magnitude under a $\pm$1.8 GeV variation of \mt,
as shown in Figure~\ref{fig:intro}(b), from Ref.~\cite{Degrassi}).

Currently, the most precise measurements of \mt\ are obtained from
direct reconstruction of the top-quark decay final states and use
calibrations based on Monte Carlo (MC) simulation to determine the
top-quark mass value that best describes the data. In this approach,
the measured top-quark mass corresponds to the parameter implemented
in the MC (\mtmc) which formally is not a renormalised field theory
parameter, and must be used with care as input for precise theoretical
predictions~\cite{Mitov,Hoang,Weinzierl}. The top quark is colour
charged and does not exist as an asymptotic state: the value of \mt,
extracted from the experiments, depends on the theoretical definition
of the mass, which varies according to the renormalisation scheme
adopted: pole mass (\mtpole) or running mass.  As a result, the
identification of \mtmc\ with \mtpole\ is currently subject to an
uncertainty of the order of 1 GeV~\cite{Hoang}, comparable to the
present experimental precision (see also Refs.~\cite{mtr2013, mitp2014}
for previous recent reviews on \mt).

\section{Top-quark pair production and signatures at the Tevatron and
  LHC}

At Tevatron and LHC hadron colliders top quarks are mainly produced in
pairs, through strong interactions, via gluon fusion and
quark-antiquark annihilation processes. Depending on the collider
centre-of-mass-energy ($\sqrt{s}$), and on the type of particle beams
being utilised (proton-antiproton, $p\bar p$, or proton-proton, $pp$),
the relative importance of the two processes varies. At the Tevatron
$p\bar p$ collider, operating at $\sqrt{s}=1.8-1.96$~TeV,
approximately 85\% of the top-quark pairs (\ttbar) are produced
through quark-antiquark annihilation, whereas at all
centre-of-mass-energies explored by the LHC $pp$ collider, gluon
fusion processes are dominant (80$-–$90\% for $\sqrt{s}=7-–14$
TeV).
The top-quark pair production cross-section varies from
$7.16^{+0.20}_{-0.23}$~pb at the Tevatron, to $172.0^{+6.4}_{-7.5}$~pb
($\sqrt{s}=7$~TeV), $245.8^{+8.8}_{-10.6}$~pb ($\sqrt{s}=8$~TeV) and
$953.6^{+27.9}_{-38.3}$~pb ($\sqrt{s}=14$~TeV) at the
LHC~\cite{Czakon}.
The production of single-top quarks occurs via electroweak
interactions and relates to a significantly lower (about one half)
production cross-section than that for top-quark pairs.

After the discovery of the top quark in
1995~\cite{topobsCDF,topobsDZE}, the CDF and D0 experiments, operating
at the Tevatron, have collected about 10~\ifb\ of $\sqrt{s}=1.96$~TeV
$p\bar p$ collisions.  The LHC experiments, ATLAS and CMS, in
operation since 2010, have collected 5~\ifb\ and 20~\ifb\ of $pp$
collisions data at the centre-of-mass-energies of 7 and 8~TeV,
respectively (LHC Run-1). Within the planned LHC programme, about
100~\ifb\ of $\sqrt{s}=13–-14$~TeV (LHC Run-2) and 200~\ifb\ of
$\sqrt{s}=14$~TeV (LHC Run-3) $pp$ collision data will be collected in
the time period 2015$–-$2022.  An additional ten-fold increase of
the integrated luminosity is expected within the LHC high-luminosity
upgrade~\cite{HLLHC}.
Correspondingly, the expected number of \ttbar\ events that will be
produced by the end of LHC Run-3 amount to about 300 Million, compared
to 6 Million produced during the LHC operations at $\sqrt{s}=7$ and 8
TeV, and 70k produced at the Tevatron at $\sqrt{s}=1.96$~TeV. As we
shall see in the following, this will open unprecedented opportunities
for precise measurements of the properties of the top quark, and in
particular of \mt.

In the SM, the top quark decays almost exclusively into a $W$ boson
and a $b$-quark ($t\to Wb$). The \ttbar\ signatures are therefore
determined by the $W$ boson decay modes. In the ``all-jets'' channel,
with a branching ratio (BR) of 46\%, both $W$ bosons decay into a
quark-antiquark pair ($W\to q\bar q'$). In the ``lepton+jets'' channel
(BR=44\%), one $W$ boson from the top or antitop quark decays to a
pair of charged and neutral leptons ($e, \mu, \tau$, and their
corresponding neutrinos), while the other $W$ boson decays into a
quark-antiquark pair.  Finally, the \ttbar\ ``di-lepton'' channel
corresponds to the case where both $W$ bosons from the top and
anti-top quarks decay leptonically, into a pair of charged and neutral
leptons (BR=10\%)~\cite{PDG2014}.
The experimental signatures associated to single-top quark production
vary depending on the production mode ($s$-, $t$- and $tW$-channels).
Currently, \mt\ measurements are available only for the $t$-channel,
typically characterised by the presence of a detectable jet recoiling
against the produced $t$ (or $\bar t$) quark. To facilitate the event
reconstruction and to distinguish the single-top quark production from
the overwhelming background processes, the $W$ boson associated to the
top quark is required to decay leptonically ($W\to l\nu$).

\section{Experimental setups, event selection and reconstruction}

The CDF~\cite{CDF,CDFbis}, D0~\cite{DZE,DZEbis}, ATLAS~\cite{ATL} and
CMS~\cite{CMS} experiments utilise omnipurpose detectors that have
been designed for the identification and reconstruction of the
particles emerging from $p\bar p$ or $pp$ collisions provided by the
Tevatron and LHC accelerator complexes. Despite the different
underlying technological choices, all detectors comprise three major
subsystems.
In the region close to the interaction point, tracking systems
immersed in a magnetic field record precisely the trajectories and
transverse momenta of charged particles. The energy and position of
the electromagnetic and hadronic showers are measured by means of
hermetic and finely grained calorimeter systems located immediately
following the tracking systems. Finally, the outer part of the
detectors comprises of dedicated muon systems providing precise
momentum measurements for highly penetrating and energetic particles.

In general, the design and calibration of experimental physics
analyses proceed via the use of MC simulated signal (\ttbar\ or
single-top quark) and background events. The generation of a primary
hard interaction process ({\it e.g.}  $q\bar q, gg\to \ttbar$), is
accompanied by parton showers, and by non-perturbative processes that
convert the obtained partonic final state into colourless hadrons.
Subsequently, soft interactions compounding on the event of interest
(``pile-up''\footnote{Pile-up is the term given to the extra signal
  produced in the detector by $\ppbar$ or $pp$ interactions other than
  the primary hard scattering.}), are also included in the
simulation~\cite{MassDef}. The MC events are processed through
experiment-specific simulation and reconstruction software, and the
reconstructed final-state particles, originating from quarks and
gluons evolving into collimated sprays of colourless hadrons, are
clustered into jets, that can be associated with the final-state
partons.

Key ingredients for \mt\ analyses are the jet energy scale (JES)
calibration procedures. These are applied after jet reconstruction,
are meant to ensure the correct measurement of the average jet energy
across the whole detector (inter-calibration), and are designed to be
independent of the pile-up conditions.
Jet energy corrections account for the energy lost in un-instrumented
regions between calorimeter modules, for differences between
electromagnetically and hadronically interacting particles, as well as
for calorimeter module irregularities.
The calibration procedures use single hadron calorimeter response
measurements, systematic MC simulation variations as well as in-situ
techniques, where the jet transverse momentum ($p_{\rm T}$) is
compared to the $p_{\rm T}$ of a reference object, for example in
$\gamma$+jets and $Z$+jets events~\cite{JET-ENERGY-SCALE-CDF,
  JET-ENERGY-SCALE-DZERO-NEW, ATLASJESPAPER2015, ATLASJESPAPER2010,
  CMSJES, CMSJES2}.
Uncertainties on the JES vary from about 1\% to 3\% depending on the
jet kinematic properties and flavour ($u,d,c,s,b$ or gluon originated
jets), and are typically among the largest sources of systematic
uncertainty in \mt\ measurements.

\subsection{Event selection}
Event selection requirements, targeted at the $t\bar t$ (or $t$)
signature under study, rely on the number of well-reconstructed
physics objects in the detector, jets and charged leptons (typically
electrons and muons), and on their properties. The presence of
identified $b$-quark jets (by means of $b$-tagging algorithms, based
on the properties of secondary vertexes and low $p_{\rm T}$ leptons
reconstructed within the
jets~\cite{CDFBTAG,D0BTAG,ATLASBTAG,ATLASBTAG2,ATLASBTAG3,CMSBTAG}),
and of significant missing transverse energy (\met) from the
undetected neutrino(s), are in general also required. The experimental
signature, referred to as ``\met+jets'' channel, selects events solely
based on \met\ and jet related information rather than on explicit
charged lepton identification criteria thus achieving high acceptance
to $W\to \tau \nu$ decays.

\subsection{Event reconstruction}
\label{sec:reco}
In the \ttbarlj\ channel, selected events are typically subject to a
kinematic fit aimed at extracting the full information from the
underlying \ttbar\ decay.  Despite implementation differences, the
available algorithms, $p_{\rm T}$-max, $\chi^2$- or likelihood-based
(see Ref.~\cite{KLFitter} for a comparison of their performance),
relate the measured kinematics of the reconstructed objects to the
leading-order representation of \ttbar\ decay, and return the best
jet-to-parton association, according to the experimental resolution,
to be used in the analyses for example in the calculation of the
top-quark and $W$-boson invariant masses.  Missing information,
related to the longitudinal boost of the escaping neutrino(s), can be
recovered by additional constraints to the event kinematics:
exploiting the known $W$-boson mass (and requiring $m_{q\bar q^\prime}
= m_{l\nu} = m_W$) and by imposing the top- and antitop-quark masses
to be equal.
The general idea of the kinematic fit is extended to events with no
reconstructed charged lepton in the \ttbaraj\ and \ttbarmet\ channels
(for the latter, the leptonically decaying $W$ boson is treated as
missing particle as a whole).

In the \ttbarll\ channel, due to the presence of two escaping
neutrinos, the full kinematic configuration of the event cannot be
resolved unless additional assumptions are made.
A possible solution is to construct \mt-sensitive observables only
using the available information (neglecting the presence of
neutrinos), and measure \mt\ based on the kinematics of the
identified ($b$-)jets and charged leptons in the event. For example
\mt\ can be determined exploiting its correlation to the invariant
mass of the charged lepton and $b$-jet system, $m_{lb}$.  A similar
approach can be followed for the \mt\ analyses based on single-top
quark signatures.
Alternatively, more elaborate methods, {\it e.g.} ``neutrino
weighting''~\cite{NuWeight1, NuWeight2} or ``analytical matrix
weighting''~\cite{CMSDL7} techniques, targeted at obtaining
information about the complete $\ttbar$ kinematics, can be exploited.
The neutrino weighting approach steps through different hypotheses for
the pseudo-rapidity of the two neutrinos in the final state, and for
the underlying \mt.  For each hypothesis, the algorithm calculates the
full event kinematics and assigns a weight to the resulting
reconstructed top-quark mass based on the agreement between the
calculated and measured \met. The solution corresponding to the
maximum weight is selected to represent the event and the underlying
$\ttbar$ decay.
Similarly to the neutrino weighting case, in the analytical
matrix weighting technique the full reconstruction of the event
kinematics is done under different \mt\ assumptions. For each event,
the most likely \mt\ hypothesis, fulfilling \ttbar\ kinematic
constraints, is obtained by assigning weights that are based on
probability density functions for the energy of the charged leptons
taken from simulation, which are applied in the solution of the
kinematic equations~\cite{CMSDL7}.

\begin{figure}[h!]
\centering
{\includegraphics[width=.95\textwidth]{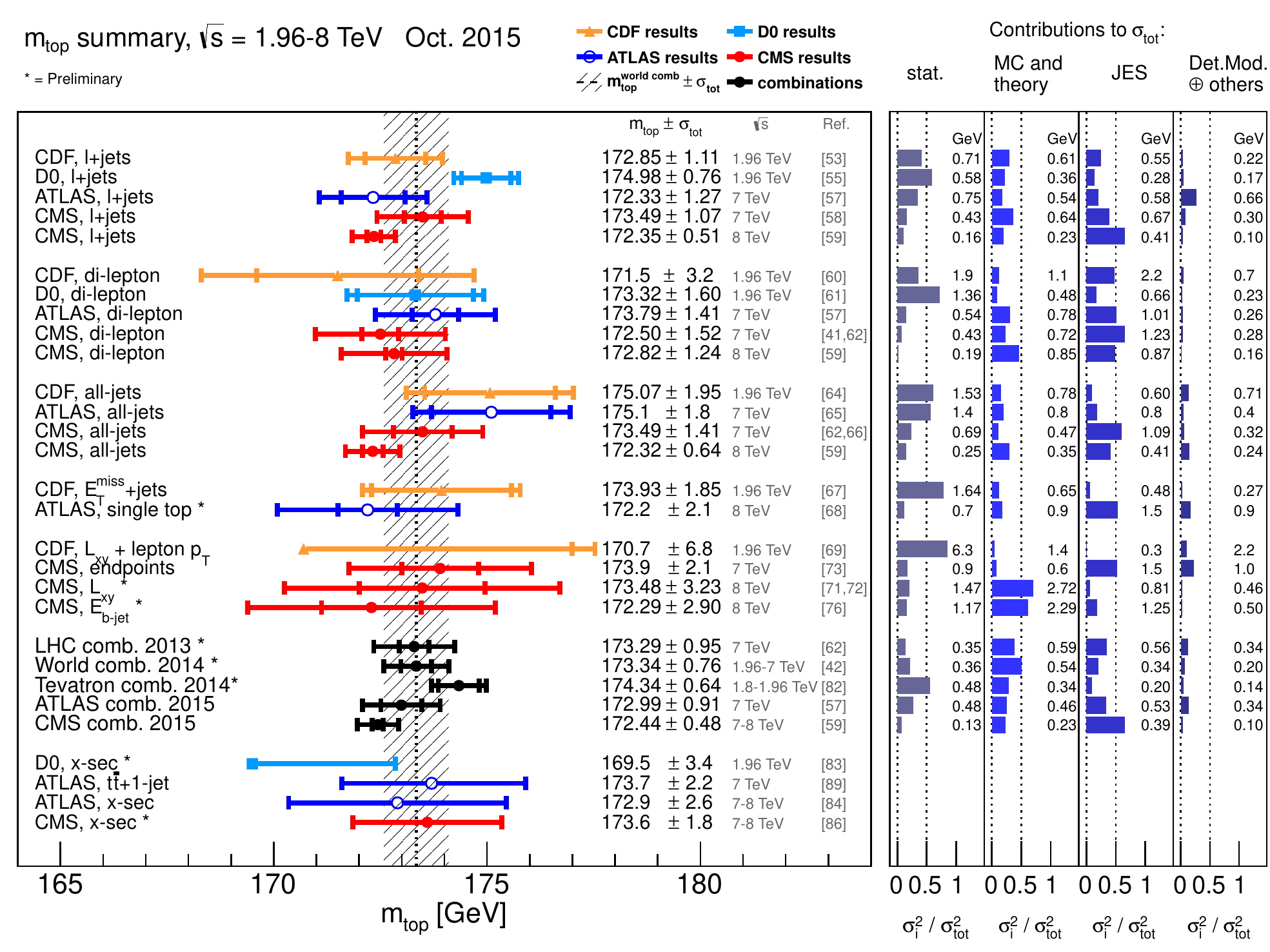}}

\caption{(left panel) Summary of the latest \mt\ measurements and
  combinations. The total \mt\ uncertainty (in all cases symmetrised),
  the relative importance (defined as $\sigma_i^2/\sigma_{\rm tot}^2$) and the size (in GeV) of
the main uncertainty contributions are provided (right panel). 
All measurements are compared to the result
of the world (Tevatron+LHC) combination~\cite{WA2014}.
For the \mt\ results
based on $L_{xy}$ and
lepton $p_{\rm T}$ (CDF) and on the inclusive \ttbar\ production cross
section (D0) only the upper error bars are shown.
\label{fig:summary}}
\end{figure}

\section{Overview of the \mt\ measurements}

The top-quark mass is measured using various techniques and in
different decays channels by the CDF, D0, ATLAS and CMS experiments.
The latest and most precise results are summarised in
Figure~\ref{fig:summary}, and compared to the results of the first
Tevatron+LHC \mt\ combination~\cite{WA2014}. The measured \mt\ values
(left panel), their total uncertainty and the (relative) importance of
the main uncertainty categories (right panel) are provided.
Two main classes of measurements can be distinguished:
\begin{itemize}
\item{} Direct \mt\ measurements, exploiting information from the
  kinematic reconstruction of the measured top-quark decay
  products, and their corresponding combinations (summarised in this Section).
\item{}  Indirect determinations of \mt, based on the comparison of inclusive or
  differential \ttbar\ production cross-section to the corresponding theory
  calculations (detailed in Section~\ref{sec:alternativetechniques}).
\end{itemize}

\subsection{Methodology}
\label{sec:techniques}

The methods exploited for the measurement of \mt\ directly using the
kinematic properties of the \ttbar\ (or single-top quark) decay
products can be categorised in the following groups.

\subsubsection{Template method}

In the ``template method'', based on a full (\ttbarlj, \ttbaraj, and
\ttbarmet) or partial (\ttbarll, and single-top quark) reconstruction
of the kinematics underlying the top-quark(s) decay, probability
density functions (templates) for observables sensitive to the
underlying \mt, and to additional parameters, are constructed based on
MC simulation.
As an example, the templates may be the distribution of the top-quark
masses reconstructed from a kinematic fit in MC samples generated
using different input \mt.  The templates can be subsequently mapped
to continuous functions of \mt, either through a non-parametric
kernel-density estimator~\cite{KDE}, or by fitting an analytic
function that interpolates between the discrete input values of \mt,
and then used in a maximum-likelihood fit to the data.
Depending on the number of input distributions utilised,
one-, two- or multi-dimensional template fits are performed to
determine \mt\ along with possible additional parameters, {\it e.g.} a
global jet-energy scale factor (JSF), targeted at reducing the impact
of the experimental systematics stemming from the
JES~\cite{CDFlj2006}.
The template method, originally exploited for the \mt\ measurements in
the top-quark observation papers~\cite{topobsCDF,topobsDZE}, is widely
used for $\mt$ analyses, and constitutes the reference technique for
the latest CDF and ATLAS \mt\ results.

\subsubsection{Matrix element method}

The ``matrix element method'' is based on the likelihood to observe a
sample of selected events in the detector.
For each event a probability is calculated as a function of the
assumed values of each parameter to be measured ({\em e.g.} \mt\ and
JSF), using a leading-order matrix element incorporating the
differential cross-sections of the physics processes relevant to the
analysis.
All possible assignments of reconstructed jets to final-state partons
are used, each weighted by a probability determined from the matrix
element. The correspondence between measured four-vectors and
parton-level four-vectors is taken into account using probabilistic
transfer functions (see Ref.~\cite{MEM} for a comprehensive review of
the method).
This approach maximises the statistical power of the considered data
sample at the cost, however, of a high computational demand. As a
consequence, this technique is best suited for small data sets, for
example for event samples obtained by means of very tight selection
criteria, or exploiting exclusive \ttbar\ decays with reduced
branching ratios.

\subsubsection{Ideogram method}

The ``ideogram method''~\cite{IdeogramMeth} combines some of the
features of the above two techniques, and can be considered a
computational effective approximation of the matrix element method.
After a kinematic fit of the decay products to a \ttbar\ hypothesis
(Section~\ref{sec:reco}), MC-based likelihood functions are exploited
for each event (ideograms) that depend only on the parameters to be
determined from the data. The ideograms reflect the compatibility of
the kinematics of the event with a given decay hypothesis.
As in the case of the template and matrix element methods, ideograms
can be generalised in multiple dimensions depending on the number of
input observables used.

\subsection{Uncertainties affecting the \mt\ measurements}

\subsubsection{Statistical uncertainties}
Statistical uncertainties on \mt\ arise from the finite size of the
data samples available for the measurements.  For Tevatron and LHC
Run-1 analyses the statistical uncertainty constitutes a sizable
contribution to the total uncertainty, especially for measurements
exploiting multidimensional fits to the data, to determine
simultaneously \mt\ and additional parameters designed to reduce the
impact of the JES uncertainty on the measurement  (see
Sections~\ref{sec:detmodunc} and \ref{sec:results})\footnote{In the
  present work, the full statistical uncertainty of the fit is quoted,
  including the contributions from the simultaneous determination of
  additional parameters along with \mt.}.  The statistical
uncertainty component is expected to be reduced by up to one order of
magnitude for the upcoming LHC analyses exploiting the full Run-2,3
data.

\subsubsection{MC/theory related uncertainties
(MC/theory)}
Theory based uncertainties are related to the simulation of top-quark
pair (or single-top quark) signal events, to the event modelling and
to the description of the hard scattering environment.  Choices to be
made in the simulation are the proton (antiproton) parton distribution
functions (PDFs), the MC generator and the hadronisation model. On the
event modelling side, important ingredients are the description of the
underlying event (UE) via MC tunes, and the settings adopted for the
modelling of colour re-connection (CR), extra initial- or final-state
QCD radiation (ISR/FSR) and the description of additional interactions
accompanying the hard scatter (pile-up).

Whenever possible, MC modelling uncertainties are constrained using
data. At the Tevatron, the MC parameters excursion used to evaluate
the impact of ISR/FSR effects on \mt\ is determined in Drell-Yan
events, which share the same initial state as most of the \ttbar\
pairs, by studying the kinematic properties of the di-lepton
pairs~\cite{ISR-FSR-CDF, DZELJPRD}. Constraints to the
ISR/FSR systematics are obtained at the LHC by exploiting
``gap-fraction'' (``jet-veto'') observables and the properties of extra
jets accompanying the \ttbar\ system~\cite{LHCISR,LHCISR2}.
Additional studies, currently statistically limited, based on
jet-shapes~\cite{ATLJetShapes}, or on the study of the UE and CR
kinematics in \ttbar\ events~\cite{CMSJpsi} will become more sensitive
and eventually be able to discriminate different MC models. These
investigations are expected to play a crucial role in improving the
MC/theory uncertainties in \mt\ analyses exploiting the data from the
upcoming LHC runs.

\subsubsection{Experimental uncertainties: Jet Energy Scale and
Detector Modelling (JES and Det. Mod.)} 
\label{sec:detmodunc}
Experimental uncertainties stem from the modelling of the physics
objects used for the event reconstruction and from the description of
the detector response. These are related to the identification,
reconstruction and calibration of charged leptons, jets, and \met.
The main systematic uncertainty contributions to the measurements
typically originate from the jet energy scale for light-quark ($u, d,
c, s$ or gluon) and $b$-quark originated jets (JES and bJES), and from
the uncertainty related to the modelling of the performance of the
$b$-tagging  algorithms in data relative to the MC.
Additional detector modelling uncertainties, including trigger
efficiencies, uncertainties on the data to MC modelling of the charged
leptons identification, reconstruction and energy scale, as well as
uncertainties stemming from the \met\ and pile-up effects, are
typically sub-dominant.
Profiting from the large \ttbar-enriched data sets that have become
available, the potentially large \mt\ systematics from detector
modelling effects is substantially mitigated by the application of
analysis techniques in which \mt\ is determined simultaneously with
additional in-situ parameters exploiting information of jet energy
scale sensitive distributions (Section~\ref{sec:results}). For
example, a global jet energy scale factor (JSF), defined as a
multiplicative factor to be applied in addition to the standard jet
energy corrections, can be constrained by the response of light-quark
jets using the kinematic information in $W\to q\bar q'$ decays
(referred to as in-situ \ttbar\ jet energy calibration).

Finally, uncertainties on the background normalisation and
differential distributions can affect the measured top-quark
properties. For the analyses in the lepton+jets and di-lepton
channels, in particular at the LHC, these uncertainties contribute
only marginally to the total uncertainty of \mt.

\subsection{Individual results}
\label{sec:results}
In the following, individual \mt\ results are summarised and presented
according to the final state exploited.

\subsubsection{\ttbarlj\ channel}
The lepton+jets channel yields the most precise \mt\ measurements
across all experiments. This final state profits from a good signal to
background ratio, and the possibility to fully reconstruct the event
kinematics, despite the presence of one neutrino from the $W$ boson
decay ($W\to l\nu$). Although different techniques (template,
ideograms or matrix element, see Section~\ref{sec:techniques}) are
applied to measuring the top-quark mass, all recent analyses mitigate
the systematic uncertainty due to JES uncertainty by a simultaneous
in-situ fit to global jet energy scale factor (JSF) sensitive
distributions.

The CDF \ttbarlj\ analysis~\cite{CDFLJ} is based on the template
method, and uses 8.7~\ifb\ of $p\bar p$ collisions at
$\sqrt{s}=1.96$~TeV. Events are reconstructed by means of a
$\chi^2$-based kinematic fit which determines the best jet-to-parton
assignments. To increase the statistical power of the analysis, two
\mt-related invariant masses ($\mt^{\rm reco}, \mt^{\rm reco2}$),
corresponding to the best and second best jet-to-parton assignments
and the invariant mass of the two jets from the hadronically decaying
$W$ boson ($m_{q\bar q^\prime}$, sensitive to JSF) are fit to the
data. The jet energy calibrations are improved using an artificial
neural network to achieve a better $b$-jet energy resolution. In a way
similar to what is described in Ref.~\cite{CDFJetNNCorr}, this
algorithm incorporates precision tracking and secondary vertex
information, in addition to standard calorimeter measurements. The
final \mt\ result, $\mt = 172.85\pm 1.11$~GeV, is obtained from a
simultaneous fit to five event sub-samples defined according to the
$b$-tagged jet multiplicity and properties, and is determined along
with a JSF via a three-dimensional template method.  The overall \mt\
uncertainty receives comparable contributions from the
statistical, the MC/theory and JES uncertainties
(Figure~\ref{fig:summary}).

The D0 \ttbarlj\ analysis~\cite{DZELJPRD, DZELJ} rests on the matrix
element technique and utilises the full $p\bar p$ data set at
$\sqrt{s}=1.96$~TeV provided by the Tevatron, corresponding to an
integrated luminosity ($\mathcal{L}$) of 9.7~\ifb. The analysis
profits from updated JES
calibrations~\cite{JET-ENERGY-SCALE-DZERO-NEW} and an improved
implementation of the matrix element method~\cite{AcceleratingMEM}.
The latter reduces the computational demand by two orders of
magnitude, allowing a substantial increase of the number of simulated
MC events used for the method calibration, and the evaluation of
systematic uncertainties.  A simultaneous fit to the data is performed
to obtain \mt\ and JSF, and results in the most precise Tevatron \mt\
measurement: $\mt = 174.98 \pm 0.76$~GeV, where the largest
contribution to the total uncertainty is statistical in nature
(0.58~GeV).
 
A recent extension of the template method in the \ttbarlj\ channel as
been proposed and exploited by ATLAS, where \mt\ is determined
simultaneously with a JSF from $W\to q\bar q'$ decays and a separate
$b$-to-light-quark energy scale factor (bJSF)~\cite{ATLLJDL7}. The
bJSF sensitive observable, $R_{bq}$, is defined in terms of a ratio of
the scalar sums of the $p_{\rm T}$ of the $b$-tagged and light-quark
jets present in the event.
The JSF and bJSF account for differences between data and simulation
in the light-quark and in the relative $b$-to-light-quark jet energy
scale, thereby mitigating the corresponding systematic
uncertainties. The result, $\mt = 172.33 \pm 1.27$~GeV, is based on
4.6~\ifb\ of LHC $pp$ collision data at $\sqrt{s}=7$~TeV, and has a
sizable contribution from the statistical uncertainty (0.75~GeV), due
to the dimensionality of the fit. This is expected to be greatly
reduced when applying the method to the four-fold larger 8~TeV data
set.

The CMS collaboration reports \mt\ measurements in the \ttbarlj\
channel based on the full LHC Run-1 data sets ($\mathcal{L} =
5.0~\ifb$ and $\mathcal{L} = 19.7~\ifb$) collected at $\sqrt{s}=7$~TeV
and 8 TeV~\cite{CMSLJ7, CMS8}. In both analyses, \mt\ is derived
simultaneously with a JSF from $t\to Wb~(W\to q\bar q')$ decays,
employing a two-dimensional ideogram method. While in the 7~TeV
analysis a simultaneous fit to \mt\ and JSF is performed assuming no
prior knowledge of the JSF, a JSF constrained fit is applied to the 8
TeV data set (referred to as ``hybrid'' method in Ref.~\cite{CMS8}).
This procedure incorporates the prior knowledge about the JES and its
uncertainty, applying a Gaussian constraint to JSF.
The hybrid approach is found to reduce both the statistical and
systematic uncertainties compared to the unconstrained two-dimensional
fit, yielding a total uncertainty improvement of about 20\%. The
measured \mt\ values are: $\mt = 173.49 \pm 1.07$~GeV and $\mt =
172.35 \pm 0.51$~GeV for the 7 and 8~TeV analyses, respectively, where
the overall uncertainties are driven by residual JES and MC/theory
based systematic uncertainties.  The CMS 8~TeV \ttbarlj\ result
constitutes the most precise \mt\ measurement to date.

\subsubsection{\ttbarll\ channel}

In the di-lepton channel, the signal to background ratio is typically
very good, and the kinematics is under constrained due to the presence
of at least two neutrinos in the final state. As a result, direct
in-situ fits to global jet energy scale factors are inhibited and the
corresponding JES uncertainties are larger than for the lepton+jets
results (the JES contribution accounts typically for 50\%--60\% of the
total uncertainty on \mt, Figure~\ref{fig:summary}).
In addition, due to the relatively small branching ratio, Tevatron
\mt\ analyses exploiting this decay mode are characterised by
fairly large statistical uncertainties (1.4--1.9~GeV).

The CDF \ttbarll\ analysis~\cite{CDFDL} uses 9.1~\ifb\ of 1.96 TeV
$p\bar p$ collisions, and exploits templates of a special observable,
$m^{\rm hyb}$, for an optimal reduction of the dominant JES systematic
uncertainty. The $m^{\rm hyb}$ is constructed as a linear combination
of the top-quark mass obtained from the neutrino-weighting algorithm
and $m_{lb}^{\rm alt}$, a variable defined based on the ratio of the
invariant masses of the lepton and $b$-jet systems, and the $b$-jets
energies, to minimise the sensitivity to the JES. The result yields
$\mt = 171.5\pm 3.2$~GeV, where the total uncertainty receives
comparable contributions from the statistical (1.9~GeV) and systematic
(2.5~GeV) uncertainties.

The latest \ttbarll\ D0 result~\cite{DZEDL}, is based on 9.6~\ifb of
$p\bar p$ collisions at $\sqrt{s}=1.96$ TeV. It features a
comprehensive optimisation of the neutrino weighting method and
fitting parameters to minimise the statistical uncertainties, and
profits from a JSF recalibration based on the results the \ttbarlj\
analysis~\cite{DZELJ}, which reduces the otherwise limiting JES
systematic uncertainty. The observables used to determine \mt\ via a
two-dimensional template method are the first moments, mean and
standard deviation, of the $\mt^{\rm reco}$ distributions obtained
from the neutrino weighting algorithm. The resulting $\mt = 173.32 \pm
1.60$~GeV constitutes the best di-lepton result from the Tevatron and
is competitive with the results from the LHC.

The ATLAS $\ttbarll$ analysis is based on a one-dimensional template
method. Instead of attempting a full kinematic reconstruction,
templates are obtained for the $m_{lb}$ observable, defined as the
per-event average invariant mass of the two lepton (either electron or
muon) plus $b$-jet pairs from the decay of the top
quarks~\cite{ATLLJDL7}. In contrast to D0, to keep the correlation to
the $\ttbarlj$ result minimal, thereby maximising the gain in the
ATLAS \mt\ combination, the jet energy scale factors (JSF, bJSF)
measured in the $\ttbarlj$ channel are not propagated to di-lepton analysis.
The final result, $\mt = 173.79\pm 1.41$~GeV, is obtained from a
simultaneous fit to the sub-samples of events defined according to the
$b$-tagged jet multiplicity, and has an overall correlation of
$\rho_{\rm tot}=-7\%$ to \ttbarlj\ \mt\ result.

Measurements of \mt\ in the di-lepton channel are available from the
CMS collaboration based on the LHC Run-1 data sets at
$\sqrt{s}=7$~TeV~\cite{CMSDL7, LHC2013} and 8~TeV~\cite{CMS8}
corresponding to integrated luminosities of 5.0~\ifb\ and 19.7~\ifb,
respectively. The top-quark mass is reconstructed with the analytical
matrix weighting technique, and the measurements use distributions
derived from MC signal samples, generated at different input \mt\
values, and backgrounds, in binned likelihood fits to the data. While
different templates are used according to the $b$-tagged jet
multiplicity of the event in the 7~TeV analysis, as a result of an
optimisation procedure to minimise the total uncertainty, only events
with two $b$-tagged jets are used for the 8~TeV results. The resulting
top-quark masses are $\mt = 172.50 \pm 1.52$~GeV and $\mt = 172.82 \pm
1.24$~GeV for the 7 and 8~TeV analyses, respectively\footnote{In
  addition, an alternative \mt\ result in the \ttbarll\ channel is
  presented in Ref.~\cite{CMSmlb} based on the $\sqrt{s}=8$~TeV data
  set and exploiting the $m_{lb}$ observable ($\mt = 172.3\pm
  1.3$~GeV). The analysis setup is also used to extract the top-quark
  mass using different theory predictions, MC simulations as well as a
  fixed-order QCD calculations.}.

\subsubsection{\ttbaraj\ channel}

In the all-jets channel a full reconstruction of the event kinematics
is possible without the ambiguity due to neutrino momenta, however the
signal to background ratio is significantly poorer due to the severe
QCD multijets background, whose production cross-section exceeds that
of \ttbar\ by several orders of magnitude.
Despite this underlying limitation, and the particular attention
required to precisely estimate and control the background
contributions via data-driven techniques, the final \mt\ precision
obtained in this channel is comparable to that of the lepton+jets and
di-lepton results.

CDF measures \mt\ in the \ttbaraj\ channel using 9.3~\ifb\ of $p\bar
p$ collision data at $\sqrt{s}=1.96$~TeV~\cite{CDFAJ}.  To strongly
suppress the background, the event selection is complemented by a
multivariate algorithm, containing multiple kinematic and jet-shape
variables as input.
Using a two-dimensional template method, a top-quark mass of $\mt =
175.07 \pm 1.95$~GeV is measured along with a JSF exploiting
information from the hadronically decaying $W$ boson, and
incorporating a prior based on the external JES uncertainty.

With the ATLAS \ttbaraj\ analysis based on 3.5~\ifb\ of $pp$
collisions at $\sqrt{s}=7$~TeV~\cite{ATLAJ7}, the top-quark mass is
obtained from a one-dimensional template fit to the ratio of three-jet
to di-jet mass ($R_{3/2}= m_{jjj}/m_{jj}=m_{q\bar q'b}/m_{q\bar q'}$).
The three-jet mass is calculated from the jets associated via a
kinematic likelihood fitter~\cite{KLFitter} to a top-quark decay.
Using these three jets the di-jet mass is obtained from the two jets
assigned to the $W$ boson decay.  While retaining sensitivity to the
underlying \mt, the $R_{3/2}$ observable allows a cancellation of
systematic effects common to the $m_{jjj}$ and $m_{jj}$ masses, thus
minimising the impact of the JES uncertainty on \mt\ in a
complementary way with respect to a simultaneous determination of \mt\
and JSF. The measurement yields $\mt = 175.1 \pm 1.8$~GeV, and the
total uncertainty receives similar contributions from the statistical
(1.4~GeV) and systematic (1.2~GeV) uncertainties.

The CMS measurements in the \ttbaraj\ channel, using 7 and 8 TeV, are
based on the ideogram method~\cite{CMS8,LHC2013,CMSAJ7}.
Within the 7~TeV analysis only \mt\ is extracted from a fit to the
data (one-dimensional ideogram method). The analysis setup with a
simultaneous determination of \mt\ and JSF is found to be subject to
comparable total systematic uncertainties: due to the tight jet
selection criteria applied, the reduction of the JES uncertainty via a
two-dimensional ideogram method is compensated by the increased
statistical uncertainty (two versus one parameter fit), and by an
enhanced sensitivity to MC modelling effects.  A similar situation is
observed for the 8~TeV analysis~\cite{CMS8}, where, similarly to the
corresponding \ttbarlj\ case, the fit employs an hybrid method with a
constrained JSF. The resulting measured \mt\ values are: $\mt = 173.49
\pm 1.41$~GeV and $\mt = 172.32 \pm 0.64$~GeV for the 7 and 8~TeV
analyses, respectively.

\subsubsection{Alternative final states and techniques for direct \mt\ determination}

Total \mt\ uncertainties comparable with those of the analyses in the
di-lepton and all-jets channels can be achieved by exploiting
alternative final states (\met+jets or single-top quark enriched), as
well as techniques based mostly on tracking information ($L_{xy}$ and
lepton $p_{\rm T}$), or on different observables (kinematic endpoints
and $b$-jet energy spectra).

The CDF \mt\ measurement in the \met+jets channel~\cite{CDFMJ} uses
8.7~\ifb\ of $p\bar p$ collisions at $\sqrt{s}=1.96$~TeV, and focuses
on events with large \met\ and jets, vetoing identified charged
leptons.  Although no identified leptons are present, the measurement
is sensitive to all $W$-boson leptonic decays, including
$W\to\tau\nu$, which constitute approximately 40\% of the signal
sample.
After selection, events are reconstructed by means of a modified
$\chi^2$-based kinematic fit, allowing for two missing particles (the
charged lepton and the neutrino associated to the $W$ boson).
Similarly to the \ttbarlj\ result~\cite{CDFLJ}, the analysis is based
on templates of the $\mt^{\rm reco}$, $\mt^{\rm reco2}$ and $m_{q\bar
  q'}$ distributions which are fit to the data. This results in a \mt\
value of $\mt=173.93\pm 1.85$~GeV, where the main contribution
(1.64~GeV) to the total uncertainty is statistical in nature.

Using 20.3~\ifb\ of $pp$ collision data at $\sqrt{s}=8~$TeV, ATLAS
measures \mt\ based on a single-top quark enriched final
state~\cite{ATL8stop}.
Selected events, targeted at the $t$-channel, contain one charged
lepton, \met, and two jets, one of which is required to be $b$-tagged,
resulting in a statistically independent data set with respect to
other ATLAS \mt\ analyses.  In addition, the ambiguities related to
the jet-to-parton assignment are minimised in this channel, and the
sensitivity to MC modelling effects is complementary to that of $t\bar
t$-based \mt\ methods, due to the different colour connection patterns
and momentum transfer scales involved.
A one-dimensional template method is used, based
on the invariant mass of the lepton and the $b$-tagged jet as
estimator ($m_{lb}$), and yields $\mt = 172.2 \pm 2.1$~GeV, where the
total uncertainty is dominated by the uncertainties on the JES
(1.5~GeV).

The CDF collaboration, using a partial $p\bar p$ data set
($\mathcal{L} = 1.9~\ifb$) at $\sqrt{s}=1.96$~TeV, developed \mt\
analysis techniques using observables with minimal dependence on the
JES~\cite{CDFLxy}. These are based on the transverse decay length of
$b$-tagged jets ($L_{xy}$), the $p_{\rm T}$ of electrons and muons
from $W$-boson decays, or a combination of both~\cite{Lxy}. Events are
selected in the \ttbarlj\ channel, and the top-quark mass measurement
is performed through comparisons with the mean $L_{xy}$ and mean
lepton $p_{\rm T}$ from MC simulations performed for a variety of
top-quark mass hypotheses. The analysis is sensitive to different
event characteristics than typical \mt\ measurements, and requires
ad-hoc data-driven calibrations of the observables and of the boost of
the top quarks. The combination of the \mt\ results obtained by the
individual $L_{xy}$ and lepton $p_{\rm T}$ observables yields $\mt =
170.7 \pm 6.8$~GeV, where the precision is limited by the statistical
uncertainty (accounting for 6.3~GeV).

At the LHC, the $L_{xy}$ technique is exploited in the \ttbarlj\ and
\ttbarll\ channels by the CMS collaboration, and applied to the $pp$
data sets collected at $\sqrt{s} = 8$~TeV, corresponding to
$\mathcal{L}= 19.3-19.6~\ifb$~\cite{CMSLxy, CMSLxy2}.  The \mt\ is
obtained using the median of the $L_{xy}$ distribution reconstructed
in data, compared to the result of MC simulations performed at
different input \mt. The result is $\mt = 173.48\pm 3.23$~GeV and the
achieved precision is limited by the uncertainties in the modelling of
the $p_{\rm T}$ of the top quark (2.6~GeV).  As in the case of the
corresponding CDF analysis, however, the JES uncertainty contributes
only marginally to the total uncertainty of \mt, signalling the high
level of complementarity of tracking based methods with respect to the
standard analyses described in the previous sections.

The CMS collaboration reports a measurement of \mt\ exploiting the
endpoints of kinematic distributions, based on 5.0~\ifb\ of $pp$ data
at $\sqrt{s}=7$~TeV~\cite{CMSendpoints}.  The method, originally
developed to determine possible NP particle masses in decay chains
with undetected particles and unconstrained kinematics, suits well the
case of \ttbarll\ decays and is based on the ``stransverse mass'',
$m_{T2}$~\cite{MT2}. To fully determine the di-lepton kinematic, the
two multistep $t\to Wb \to l\nu b$ decay chains are split and their
elements grouped in independent ways, either using only charged
lepton or $b$-jet information, or a combination of the two in the
form of an $m_{lb}$-like invariant mass. In a demonstrative effort,
motivated primarily by future applications to NP scenarios, in
addition to the top-quark mass, the masses of the $W$ boson and the
neutrino ($m_\nu^2$) are determined. These are however constrained
($m_\nu^2=0$ and $m_W= 80.4$~GeV) to achieve the best \mt\ precision.
The result: $\mt = 173.9 \pm 2.1$~GeV, has a limited dependence on the
MC simulation, and brings complementary information with respect to
conventional \mt\ analyses.

Finally, following a recent theoretical proposal~\cite{Franceschini},
the top-quark mass is measured by the CMS collaboration in the
\ttbarll\ channel ($ll=e\mu$) , using 19.7~\ifb\ of $pp$ collisions at
$\sqrt{s}=8$~TeV, based on the position of the peak of the energy
spectrum of the $b$-jets~\cite{CMSbjets}.
Under the hypothesis that top quarks are produced
unpolarised~\cite{ATLtoppol}, the chosen observable is independent of
the Lorentz boosts and can be related to the energy of the $b$-quark
in the rest frame of the top quark, in turn depending on \mt. After
calibration for event selection, reconstruction, and background
contamination effects, the top-quark mass is measured to be $\mt =
172.29 \pm 2.90$~GeV, where the dominant sources of systematic
uncertainty stem from the modelling of the hard scattering process
(MC/theory) and to a lesser extent from the JES\footnote{Another
  interesting and complementary proposal, not yet exploited by the
  experiments, is represented by the ``weight function''
  method~\cite{WeightFunction}. Based on the normalised energy
  distribution of the charged lepton emitted from the parent top-quark
  in the laboratory frame, the sensitivity to \mt\ is obtained via
  weight functions constructed such that their integral $I(m)$
  vanishes for $m=\mt$.}.

\subsection{Top-quark mass combinations} 
\label{sec:combi}

Individual \mt\ results resting on various techniques and \ttbar\ (or
single-top quark) decay channels, have different sensitivities to
statistical and systematic effects, and to the details of the MC
simulation (right panel of Figure~\ref{fig:summary}).
To exploit the full physics potential of the available measurements,
and to profit from their diversity and complementarity, they are
combined, thereby further increasing our knowledge on \mt.  Input to
all combinations are the individual results with a detailed breakdown
of the uncertainties as well as the assumed correlations between
individual sources. The tasks of each combination is to determine a
mapping between corresponding uncertainty sources, to understand the
correlations in each of the categories across different analyses and
experiments, and evaluate the compatibility of the input results.
Alongside with independent and experiment-specific
combinations~\cite{ATLLJDL7, CMS8}, multi-experiment working
groups\footnote{ The Tevatron Electroweak (TEV-EW-WG) and the LHC Top
  Physics (LHC-TOP-WG) working groups. More information at
  http://tevewwg.fnal.gov and
  http://twiki.cern.ch/twiki/bin/view/LHCPhysics/LHCTopWG.} are
responsible for carrying out \mt\ combinations using measurements from
different collaborations, and to provide various sets of
recommendations aimed at refining and harmonising the statistical and
systematic uncertainty treatment in current and future measurements.
The Best Linear Unbiased Estimator method (BLUE)~\cite{BLUE1, BLUE2,
  BLUERN} is used to perform the \mt\ combinations. It determines the
coefficients (weights) to be used in a linear combination of the input
measurements by minimising the total uncertainty of the combined
result.  In the algorithm both statistical and systematic
uncertainties and the measurement correlations are taken into account,
while assuming that all uncertainties are indipendent and distributed
according to Gaussian probability density functions.

A selection of the available \mt\ measurements is used in the recent
Tevatron, LHC, and Tevatron+LHC combinations.
The current LHC and Tevatron combinations yield $\mt=173.29 \pm
0.95$~GeV and $\mt = 174.34 \pm 0.64$~GeV, and correspond to a
precision improvement of 10\% and 16\% with respect to the most
precise input measurement, respectively~\cite{LHC2013, TEV2014}.  The
first Tevatron+LHC \mt\ combination (also referred to as ``world''
combination) results in $\mt = 173.34 \pm 0.76$~GeV~\cite{WA2014} and
improves the overall \mt\ precision by 28\% with respect to the most
precise input. In general, the systematic uncertainties stemming from
the JES (and bJES) and the MC/theory modelling dominate the total
uncertainties of the combined \mt\ results.
Except for the latest CMS combination resulting in $\mt = 172.44 \pm
0.48$~GeV~\cite{CMS8}, the present ATLAS, LHC, Tevatron and
Tevatron+LHC \mt\ combinations do not include all recently improved
individual measurements. Among these are the latest \ttbarll\ results
from CDF and D0~\cite{CDFDL,DZEDL} for the Tevatron combination; and
the ATLAS 7~TeV \ttbaraj~\cite{ATLAJ7} and single-top quark
results~\cite{ATL8stop} for the ATLAS combination ($\mt = 172.99\pm
0.91$~GeV~\cite{ATLLJDL7}).  Updated inputs to the LHC \mt\
combination include the final ATLAS 7~TeV~\cite{ATLLJDL7, ATLAJ7} and
CMS 8~TeV results~\cite{CMS8} as well as the ATLAS 8~TeV \mt\ result
based on single-top enriched signatures~\cite{ATL8stop}.
Finally, the conceivable inclusion of the individual
measurements~\cite{DZELJ,ATLLJDL7,CMS8,CDFDL,DZEDL,ATLAJ7,ATL8stop},
as well as possible refinements of the intra-experiments correlation
assumptions described in Ref.~\cite{ATLLJDL7} (see also
Section~\ref{sec:prospects}), are expected to result in major overall
uncertainty improvements in future world \mt\ combinations.
For example in the case of the ATLAS combination, the precision
improvement with respect to the most precise input \mt\ result is
increased from 8\%~\cite{LHC2013} to 28\%~\cite{ATLLJDL7} when taking
into account anti-correlations effects on \mt\ systematics, introduced
by the different analyses techniques (one- versus three-dimensional
templates).

\section{Alternative \mt\ measurement methods} 
\label{sec:alternativetechniques}

The standard techniques to measure the top-quark mass, as described in
the previous sections, make use of observables obtained via a
kinematic reconstruction of the top-quark decays.
As anticipated in the introduction, all measurements of this type rely
on MC simulation for their calibration, and \mtmc\ may differ by up to
$O(1~{\rm GeV})$ from the theoretically well-defined top-quark
pole-mass, \mtpole~\cite{Hoang}.  As described in the following,
alternative techniques are targeted at allowing a better theoretical
interpretation of the measured \mt, often approaching the
precisions of the standard results.

\subsection{Top-quark mass from inclusive $\ttbar$ cross-section
  measurements} 
The theoretical dependence of the \ttbar\ production cross-section
($\sigma_{\ttbar}$) on \mt\ can be exploited to extract the mass of
the top quark, by comparing the measured cross-section to the
corresponding theory calculation~\cite{Czakon}.  In this framework,
the top-quark mass can be measured unambiguously within the
renormalisation scheme adopted for the cross-section calculation ({\it
  e.g.}  \mtpole), provided that the \mtmc\ dependence introduced by
the event selection in the experimental analysis is negligible.
To date measurements of this type are obtained by the D0, ATLAS and
CMS collaborations.

The D0 collaboration reports a measurement of $\sigma_{\ttbar}$ in
9.7~\ifb\ of $p\bar p$ collisions at $\sqrt{s}=1.96$~TeV, using
\ttbarlj\ and \ttbarll\ final states~\cite{DZEmtopxsec}. The analysis
employs multivariate techniques to build efficient \ttbar\ signal
discriminants, exploiting the kinematic features of top-quark pair
events along with $b$-tagging information.  The measured
$\sigma_{\ttbar}$, determined from a combination of the lepton+jets
and di-lepton channels, has a total relative uncertainty of 7.3\% and
a relatively weak dependence on the \mtmc\ assumed for the calculation
of the \ttbar\ signal acceptance
($d\sigma_{\ttbar}/d\mtmc \approx -0.6 \%$/GeV around
$\mtmc=172.5$~GeV). Maximising a joint likelihood including the
experimental and theoretical dependencies on \mt~\cite{Czakon}, along
with their corresponding total uncertainties, the top-quark pole-mass
is found to be $\mt = 169.5^{+3.3}_{-3.4}$~GeV.

The corresponding ATLAS result~\cite{ATLmtopxsec} rests on the \ttbar\
production cross-section measurements in the di-lepton $e\mu$ channel
performed using $\mathcal{L}= 4.6$ and 20.3~\ifb\ of $\sqrt{s}=7$ and
8~TeV LHC $pp$ data. The numbers of events with one and two $b$-tagged
jets are counted and used to simultaneously determine
$\sigma_{\ttbar}$ and the efficiency to reconstruct and $b$-tag a jet
from a top-quark decay, thereby minimising the associated systematic
uncertainties.  The total relative experimental uncertainties on
$\sigma_{\ttbar}$ of 3.8\% (7~TeV) and 4.3\% (8~TeV), and the reduced
dependence of the measured cross-section on \mtmc\ ($d\sigma_{\ttbar}
/d\mtmc = -0.28 \pm 0.03 \%/$GeV), offer the possibility of performing
a relatively precise \mt\ measurement in the pole-mass scheme.
Results are obtained for each centre-of-mass energy and then combined
to yield $\mt=172.9^{+2.5}_{-2.6}$~GeV, where the total uncertainty is
dominated by uncertainties stemming from the choice of the PDFs (using
the PDF4LHC prescriptions~\cite{PDF4LHC}) and from the variation of
the factorisation and renormalisation scales used in the theoretical
calculations~\cite{Czakon}.

Using the full $pp$ data set available at $\sqrt{s}=7$ and 8~TeV,
corresponding to integrated luminosities of $\mathcal{L}=5.0~\ifb$ and
$19.7$~\ifb, CMS extracts \mt\ based on the $\sigma_{\ttbar}$
measurement in the di-lepton $e\mu$ channel~\cite{CMSmtopxsec}.  The
analysis is performed via a template fit of signal and background
contributions to multi-differential distributions related to the
$b$-jet multiplicity and the multiplicity and transverse momenta of
the jets present in the event.  The resulting $\sigma_{\ttbar}$ is
measured with a total relative uncertainty of 3.5\% and 3.8\% for the
7~TeV and 8 TeV data sets.  The experimental dependencies of
$\sigma_{\ttbar}$ on \mtmc\, around $\mtmc=172.5$~GeV, are
approximately $-0.38\%/$GeV and $-0.55\%/$GeV for the 7~TeV and 8 TeV
data sets, respectively. A weighted average of the \mt\ results
extracted from each centre-of-mass energy is performed taking into
account the correlations of the various systematic uncertainties, and
yields $\mt=173.6^{+1.7}_{-1.8}$ GeV, where the uncertainty stemming
from the PDF is evaluated based on the NNPDF3.0 set~\cite{NNPDF3}.

\subsection{Top-quark mass from differential normalised cross-section
  measurements}

One of the main disadvantages of the \mt\ extractions from the
inclusive \ttbar\ production cross-section measurements is connected
to their relatively limited precision with respect to the direct
methods. The current relative precision of the $\sigma_{t\bar t}$
measurements at the LHC (ranging from 3.5\% to 4.3\%) is limited by
``external'' uncertainties sources (the luminosity and beam energy
measurements, and the theoretical uncertainties related to the
cross-section calculations~\cite{ATLmtopxsec}), and a variation of 5\%
of the measured $\sigma_{t\bar t}$ induces a change of 1\% on the
extracted \mt.
To overcome this difficulty, a novel technique has been
proposed~\cite{ATLtt1jetTH}: it is based on the normalised production
cross-section of \ttbar\ pairs with an additional jet, differential in
the (inverse) invariant mass of the final-state jets.
The method shares the rigorous interpretation of the mass extracted
from the inclusive \ttbar\ cross-sections, with the advantage of a
greater sensitivity (up to a factor five larger) and competitiveness
relative to analyses based on the kinematic reconstruction of the
top-quark decay products. The chosen observable inherits its
sensitivity from the \mt\ dependence of gluon radiation off
top-quarks, with enhanced effects in the phase-space region relatively
close to the \ttbar+1-jet production threshold.
The current result~\cite{ATLtt1jet}, based on 4.6~\ifb\ of
$\sqrt{s}=7$~TeV $pp$ data collected with the ATLAS detector, yields
$\mt=173.7^{+2.3}_{-2.1}$~GeV, where the dominant contribution to the
total uncertainty is statistical in nature (1.5~GeV), and is expected
to be substantially reduced when extending the analysis to the 8~TeV
data set ($\mathcal{L} = 20.3$~\ifb).

\section{Prospects and future investigations}
\label{sec:prospects}

\subsection{Refinements of the detector and MC modelling}
\label{sec:refinements}
Systematic effects stemming from the jet energy measurements are among
the dominant sources of experimental uncertainty in many physics
analyses, and in particular the uncertainty on the jet energy scale
associated with jets initiated by a $b$-quark (bJES) plays a critical
role in \mt\ precision measurements.  The largest contributions to the
bJES uncertainty stems from the modelling of the fragmentation and
hadronisation of $b$-jets.
To reduce these, the LHC Run-2,3 data sets can be used to obtain
precise in-situ measurements of the $b$-fragmentation~\cite{CMSJpsi},
by exploiting the kinematic properties of charm meson candidates ({\it
  i.e.} $D_0$, $D^\pm$, $J/\psi$) within the decay products of the
$b$-quark jets associated with top quarks.
In a complementary approach, the $b$-jet energy scale can be probed by
comparing the measured jet energy to that of well calibrated reference
objects using charged-particle tracks within jets in \ttbar\
samples~\cite{ATLASJESPAPER2015}, and $Z$+$b$-jet events~\cite{CMSZb}.
In addition, several complementary measurements using \ttbar-enriched
data sets can be performed to substantially refine the performance of
different MC generators and tunes, and to mitigate the MC/theory
related systematic uncertainties affecting top quark physics
analyses. These comprise the study of UE and CR kinematics in \ttbar\
events similar to those proposed in Ref.~\cite{CMSJpsi}, and improved
constraints to the modelling of the ISR/FSR QCD radiation accompanying
the production of top quarks using \ttbar+jets (differential)
cross-section measurements~\cite{ATLttjetsXS, CMSttjetsXS}, as well as
jet-veto and jet-shape related observables~\cite{LHCISR,LHCISR2,
  ATLJetShapes}.

\subsection{Data unfolding and comparison with theoretical
  calculations}

The $m_{lb}$ observable in \ttbarll\ events, and the differential and
normalised \ttbar+1-jet cross-section as a function of the (inverse)
invariant mass of the final-state jets, can be computed theoretically
in perturbative QCD~\cite{ATLtt1jetTH, mlbTH, mlbTH2, mlbTH3}.
In a possible extension of the current
analyses~\cite{ATLLJDL7,CMSmlb,ATLtt1jet}, the corresponding data
distributions (corrected for experimental effects and background
contamination) can be compared to: (i) MC templates, associated with
\mtmc; (ii) the corresponding theory predictions, obtained using
unambiguously defined top-quark mass schemes, {\it i.e.} \mtpole.
This approach is expected to allow assessing the dependence of the
extracted \mt\ on the different theoretical assumptions and
implementations, as well as opening the possibility to determining
experimentally, by comparing the results of (i) and (ii), the
difference between \mtmc\ and \mtpole.  A substantially reduced
uncertainty in the relation of the two quantities will allow the \mt\
precision achieved experimentally to be fully exploited in theoretical
calculations, precision SM tests and NP searches.

\subsection{Exclusive top-quark decays, $t\to Wb \to l\nu +J/\psi X$,
  and top-tagging techniques}

The study of the production of \ttbar\ pairs with a $J/\psi$ in the
final state offers an alternative method of measuring the top-quark
mass~\cite{JpsiProposal}.  In top-quark pair events with a least one
leptonically decaying $W$ boson and one of the $b$-quarks hadronising
to $J/\psi$ (with the subsequent $J/\psi \to \mu^+\mu^-$ decay), \mt\
can be measured, ideally with no or negligible systematic
uncertainties stemming from the JES, by exploiting its correlation to
the invariant mass, $m(lJ/\psi)$, of the charged lepton and
$J/\psi\to \mu^+\mu^-$ system.
Sensitivity studies have been performed within the ATLAS and CMS
collaborations, and recent exploratory analyses based on 8~TeV $pp$
data establish the first steps towards these measurements (see
Refs.~\cite{CMSJpsi,ATLJpsi} and references therein).
Currently, within the limited size of the available data samples, the
kinematic properties of the $J/\psi$ candidates reconstructed in
\ttbar\ events are found to be in reasonable agreement with various
Monte Carlo predictions. With additional statistics (LHC Run-2,3),
these investigations are expected to significantly contribute to \mt\
precision measurements by either constraining systematic uncertainty
sources related to the $b$-quark hadronisation and fragmentation
modelling, or by enabling an alternative mass measurement method, with
complementary uncertainty contributions with respect to the
traditional results.

In addition, the centre-of-mass energy of the LHC gives access to a new
kinematic regime featuring highly-boosted top quarks. Initially
developed for NP physics searches, to efficiently identify large area
``top-jets'' including all top-quark decay products, top-tagging
techniques have proven a very active research field, and their present
performance are promising in view of future top-quark properties
measurements~\cite{ATLBoosted, ATLBoosted2, CMSBoosted}.
Similarly to the case of the top-quark mass determination using
single-top quark enriched topologies~\cite{ATL8stop}, a possible \mt\
result using highly boosted top quarks is expected to have
complementary sensitivities to MC modelling effects, and feature
peculiar colour connection patterns, interesting to further
understanding the relation between \mtmc\ and \mtpole.

\subsection{Indirect \mt\ determinations from flavour physics observables}

The systematic use of flavour physics observables has been proposed as
an alternative way to measure to mass of the top quark in a robust
theoretical framework~\cite{FlavPhys}. This approach exploits the
large top-quark Yukawa coupling and its related impact on quantum loop
corrections to identify flavour physics processes with enhanced
sensitivity to \mt, and infer the top-quark mass from their
measurements.  With the present data the obtained \mt\ precision is
limited and amounts to about 8~GeV. However, the projected achievable
total uncertainty on \mt, taking into account foreseeable theoretical
and experimental progresses, indicates a reach of about 2~GeV. Among
the studied observables the mass difference in the
$B_s$ system, $\Delta m_{B_s}$, and the measurement of the branching
ratio of $B_s \to \mu^+\mu^-$ dominate the present and the projected
sensitivities to the top-quark mass.  Although the proposed
methodology is unlikely to surpass the direct \mt\ determinations in
precision, the comparison of the various results and approaches will
carry essential information, further helping to disentangle the theoretical
ambiguities in the extraction of \mt.

\subsection{Future \mt\ combinations}

The combination of the results of different analyses, within the same
experiment, or across different collaborations allows to exploit the
full physics potential of the input measurements, to profit from their
diversity, complementarity and partially correlated uncertainties, to
increase the precision of \mt.
Continuing in-depth studies of the correlation of the \mt\
measurements from the various experiments, using different top-quark
signatures and collision data sets, taking into account
possible de-correlation or anti-correlation effects, are expected to
enable significant precision improvements~\cite{ATLLJDL7}.
In addition, the use of common \ttbar\ MC samples, processed through
the experiment-specific detector simulations, is planned to precisely
assess the systematic uncertainties related to the baseline MC
choices, and to evaluate the level of compatibility between different
strategies adopted to evaluate MC/theory related uncertainties. The
results of these investigations, together with improvements on the
individual measurements, will boost the understanding of inter-experiment
correlations and is expected to substantially increase the final
precision achievable from the combination of the available \mt\
results.

\section{Conclusions}

A review of the methodologies exploited for the measurements of \mt\
at the Tevatron and LHC hadron colliders has been presented, together
with a discussion of the main theoretical and experimental
uncertainties, and the prospects for their reduction in the course of
the LHC Run-2,3.
The application of complementary techniques, based either on the
direct reconstruction of the top-quark decay products, or on the
comparison of the experimental \ttbar\ (differential) cross-section
measurements with the corresponding theoretical calculations, will
allow to reduce the theoretical uncertainties related to the
identification of the \mtmc\ parameter, used for the direct
measurement calibrations, to well defined top-quark mass schemes ({\em
  e.g.} \mtpole).
This aspect, together with significant foreseeable improvements on the
precision reach of individual \mt\ results, and optimised treatments
of the measurement correlations in future multi-experiment \mt\
combinations, provide strong confidence that an overall \mt\ precision
of the order of 200--300~MeV~\cite{CMS-PAS-FTR-13-017, Snowmass2013},
comparable with the projections for the linear collider~\cite{ILCmt},
can be achieved at the LHC.
The unprecedented precision of the \mt\ results, and their unambiguous
theoretical interpretation, will dramatically influence the
predictions regarding the stability of the Higgs field and its effects
on the evolution of the Universe, as well as the sensitivity of SM
consistency tests and indirect NP searches.

\section{Acknowledgements}

The author would like to thank Martijn Mulders and Richard Nisius for
the fruitful discussions and useful suggestions to the manuscript.

\section*{References}

\bibliography{myrevip}

\end{document}